\RequirePackage{fix-cm}
\documentclass[smallextended]{svjour3}       
\smartqed  
\usepackage{amsfonts}
\usepackage[colorlinks,citecolor=blue,linkcolor=blue,urlcolor=blue]{hyperref}
\usepackage{amsmath}
\usepackage{amssymb}
\usepackage{graphicx}
\usepackage{comment}
\usepackage{bm}
\usepackage{mathrsfs}
\usepackage{subfigure}
\usepackage{color}
\usepackage{amsmath,epsfig}
\usepackage{epstopdf}

\renewcommand{\phi}{\varphi}

\newcommand{\be}{\begin{equation}}
\newcommand{\ee}{\end{equation}}
\newcommand{\bea}{\begin{eqnarray}}
\newcommand{\eea}{\end{eqnarray}}

\newcommand{\Int}{\mathbb{Z}}

\renewcommand{\phi}{\varphi}

\usepackage[colorlinks,citecolor=blue,linkcolor=blue,urlcolor=blue]{hyperref}

\begin{document}

\title{Cryptographic One-way Function Based on Boson Sampling
}

\titlerunning{Cryptographic One-way Function Based on Boson Sampling}        

\author{Georgios M. Nikolopoulos}


\institute{Georgios M. Nikolopoulos  
\at
Institute of Electronic Structure \& Laser, FORTH, P.O. Box 1385, GR-70013 Heraklion, Greece 
\and
Institut f\"ur Angewandte Physik, Technische Universit\"at Darmstadt, D-64289 Darmstadt, Germany
\\\email{nikolg@iesl.forth.gr}
}

\date{Received: date / Accepted: date}

\maketitle

\begin{abstract} 
The quest for practical cryptographic primitives that are robust against quantum computers is of vital 
importance for the field of cryptography. Among the abundance of different cryptographic primitives one may consider, one-way functions stand out as fundamental building blocks of more complex cryptographic protocols, and  they play a central role in modern asymmetric cryptography. We propose a mathematical one-way function, which relies on coarse-grained boson sampling. The evaluation and the 
inversion of the function are discussed in the context of classical and quantum computers.  
The present results suggest that the scope and power of  boson sampling may go beyond the proof of quantum supremacy, and pave the way towards  cryptographic applications.
\end{abstract}

%
%

%
%


\section{Introduction}
\label{secI}

One-way functions play a central role in modern cryptography, because they serve as building blocks for  
many cryptographic protocols \cite{book1,book2}. A one-way function (OWF) ${\cal F}$ is  ``easy" to perform, but ``hard" on the average to invert, in terms of efficiency and speed \cite{book1,book2}.  That is, there is a large gap between the amount of computational work required for the evaluation of  $y={\cal F}(x)$ for a given input $x$, and the amount of work required to find $x$ such that  $y={\cal F}(x)$, for a given $y$.  Widely used cryptosystems 
rely on OWFs for which the non-existence of inversion algorithms has never been rigorously proved, and 
quantum computers threatens their security \cite{book3,pqc1,pqc2}. 

The design of OWFs that are robust against quantum computers is currently at the focus of extensive interdisciplinary research \cite{pqc1,pqc2,Kab00,Buh01,Got01,Cur01,Kaw05,And06,NikPRA08,qpuf1,qpuf2,public_puf,We_Yang16,Chen18,Vlachou15,IoaMos14,Fuj12}.  
In this context, quantum OWFs  (QOWFs) have been proposed as alternatives to conventional mathematical OWFs \cite{Kab00,Buh01,Got01,Cur01,Kaw05,And06,NikPRA08,qpuf1,qpuf2,We_Yang16,Chen18,Vlachou15,IoaMos14,Fuj12}. 
QOWFs exploit quantum properties of physical systems, and  provide solutions to security problems  that are not addressed by quantum key distribution e.g., verification of identities, data integrity, etc. 
Many of these solutions require the same specialized infrastructure as  quantum key distribution, 
and  they suffer from the same practical limitations as the latter \cite{limitations}. An alternative approach to quantum-safe (also referred to as post-quantum)  cryptography relies on the quest for  mathematical OWFs that are hard to invert  by classical and  quantum  means \cite{pqc1,pqc2}. Such OWFs can serve as building blocks of cryptographic protocols that  
are computationally secure against both classical and quantum adversaries, and they are fully compatible with existing communication protocols and networks. 

In this work, our aim is to suggest a way of exploiting the problem of boson sampling  (BS) for the design of a mathematical OWF. The problem of BS pertains to sampling from a boson probability distribution, with the various probabilities 
associated with the modulus squared permanents of $N\times N$ submatrices of a uniformly (Haar) random $M\times M$ unitary matrix $\hat{\cal U}$ (with $M>N$) \cite{complex1,BSintro,BSreview}.  It has been shown that exact (and presumably even approximate) BS is hard for classical computers, unless there are severe highly implausible implications for  computational complexity theory \cite{complex1}. 
By contrast, BS is native to linear quantum optics, which makes it very attractive to photonic experiments \cite{BSintro,BSreview}. 
Typically,  a photonic BS device (BSD) involves a linear photonic circuit that implements $\hat{\cal U}$ 
between  $M$ input and $M$ output ports \cite{BSreview,ExpBS1,ExpBS2,ExpBS3,ExpBS4,ExpBS5,ExpBS6,ExpBS8}. 
Injecting $N$ identical photons at the input and  using coincidence photodetection at the output, one 
essentially samples directly from the underlying photon (boson) distribution. A BSD is expected to 
outperform conventional computers in large-scale BS, thereby offering a solid proof of 
``quantum supremacy" \cite{complex1,BSreview}. Unfortunately, recent studies place the 
threshold for quantum supremacy at $N\simeq 50$ bosons and $M\sim N^2$ ports, which is well beyond the 
current and near-term experimental capabilities \cite{Neville17}. 

To the best of our knowledge,  so far 
the BS has not been discussed in connection with applications 
that go quantum supremacy and simulations. The present work is the 
first attempt in the direction of cryptographic applications. 
Our analysis, together with existing results \cite{Neville17,Perm16}, suggests that  evaluation 
of the proposed OWF for a given input can be performed on a classical computer for $N\lesssim 40$, while due to the nature of the underlying problem, construction of inversion algorithms is conjectured to be practically impossible. 
Moreover,  the resources required for its  brute-force 
inversion scale exponentially with $N$, for both classical and quantum 
adversaries.

The paper is organized as follows. After the preliminaries of Sec. \ref{secII}, in Sec. \ref{secIII} 
we propose the bootstrap technique as a means for assigning measures of accuracy to any statistic estimated on coarse-grained 
BS data. The one-way function is proposed in Sec. \ref{secIV}, 
and its properties are discussed in detail.  A summary with concluding 
remarks is given in Sec. \ref{secV}.


\section{Preliminaries}
\label{secII}

We consider BS in the dilute limit (i.e., $M \sim N^2$), where the problem can be analysed accurately in the framework of ${M}\choose{N}$ collision-free $N$-boson configurations. 
Let 
\be 
{\mathbb S}=\{{\bm s}^{(0)}, {\bm s}^{(1)},\ldots\}
\ee 
be the set of all of these configurations, which have been ordered according to some publicly known rule. 
The $\kappa$th  configuration ${\bm s}^{(\kappa)}\in{\mathbb S}$, is a tuple of $N$ distinct positive integers 
\be 
{\bm s}^{(\kappa)}:=(s_{1}^{(\kappa)},\ldots,s_{N}^{(\kappa)}), 
\ee
where $s_{j}^{(\kappa)}\in[0,M)\equiv\Int_{M}$ refers to the port that is occupied by the $j$th boson. 
When $\{M,N\}$ are publicly known, the set ${\mathbb S}$ is also publicly known, while  
its size scales exponentially with $N$ as follows  
\be
|{\mathbb S}|\geq (M N^{-1})^N, 
\label{Ssize:eq}
\ee
for $M\geq N^2$. A configuration ${\bm s}^{(\kappa)}$ is uniquely identified by its position in the set $\mathbb S$ 
(i.e., its label $\kappa$), which is a positive integer that takes values in 
$\Int_{|{\mathbb S}|} :=\{0,1,\ldots,|{\mathbb S}|-1\}$.  
Hence, there is a one-to-one correspondence between the set of possible configurations ${\mathbb S}$, and the set of positive integers $\Int_{|{\mathbb S}|}$. 
Moreover, it takes 
\[
n=N\times (\lfloor \log_2(M)\rfloor+1)\] 
bits of information to describe the ports that are occupied by the $N$ bosons 
in  configuration ${\bm s}^{(\kappa)}$, where $\lfloor \log_2(M)\rfloor+1$ 
is the bit-length of the label of a  port.   

The set of boson configurations is partitioned into $d\gtrsim M$ disjoint subsets 
$\{{\mathbb B}_0, {\mathbb B}_1,\ldots,  {\mathbb B}_{d-1}\}$, which have 
 approximately the same size i.e., $|{\mathbb B}_j|\approx |{\mathbb S}|/d$. 
Throughout this work we assume that the number of bins $d$ scales polynomially with $\{M,N\}$  (i.e., $d\sim {\rm poly}(M,N)$), and it is sufficiently large so that boson-interference effects survive the binning, and they are reflected in the coarse-grained probability distribution.  These conditions are necessary in order for the binning to be useful in the framework of the present work, as well as in the design of efficient verification schemes for BSDs \cite{BinWD16,ShchPRL16,BSpattern}. In the dilute limit of BS, related studies show that  boson-interference effects can survive the binning for 
$d\gtrsim M$ \cite{BinWD16,ShchPRL16,BSpattern,NikBroPRA16}.
For the sake of concreteness, in the following  we consider the coarse graining discussed in Ref. \cite{NikBroPRA16}.

Given an input configuration ${\bm \psi}\in{\mathbb S}$, 
the probability for obtaining configuration ${\bm \phi}\in{\mathbb S}$ 
at the output, is given by 
\[
\tilde{P}({\bm \phi}|{\bm \psi};{\cal U})=|{\rm Per}(\hat{\cal U}_{{\bm \psi},{\bm \phi}})|^2, 
\]
where ${\rm Per}(\hat{\cal U}_{{\bm \psi},{\bm \phi}})$ is the permanent of an $N\times N$ submatrix of $\hat{\cal U}$, which  is determined by the  input and output boson configurations \cite{complex1,BSintro,BSreview}. 
The OWF we propose below relies on the quest for the most probable bin (MPB) of the 
coarse-grained distribution 
\bea
{\mathscr P}({\bm \psi};\hat{\cal U}):=\{P(\beta|{\bm \psi};\hat{\cal U}):\beta\in \Int_d\},
\label{P_cgbs:eq}
\eea
where $P(\beta|{\bm \psi};\hat{\cal U})$ is the conditional probability for the bin with label $\beta$ to occur, 
given the input configuration ${\bm \psi}$ and the unitary $\hat{\cal U}$.  
Let $\mu\in\Int_d$  denote the label of the MPB ${\mathbb B}_\mu$, with the corresponding  maximum probability given by 
 \[ P_{\max}:=P(\mu|{\bm \psi};\hat{\cal U}) = \sum_{{\bm \varphi}\in \mathbb B_{\mu}}  \tilde{P}({\bm \phi}|{\bm \psi};\hat{\cal U}).\] 
For a given unitary,  both of $\mu$ and 
$P_{\rm max} $ depend strongly on the input configuration 
${\bm \psi}$ \cite{NikBroPRA16}, and cannot be predicted without knowledge of ${\bm \psi}$. 
When ${\bm \psi}$ is known,  one can sample directly from the  coarse-grained  distribution, and 
the various probabilities 
can be approximated by the frequency of occurrence of the bins in the sample, thereby obtaining an estimate of $P_{\max}$,  say $p_{\max}$, and making an inference on the label of the MPB. 

Irrespective of whether the sampling is performed by classical or quantum means, in practice one can aim only at attaining a 
$100(1-\gamma)\%$ confidence interval (CI) to $P_{\max}$,  
for some uncertainty $\gamma\ll1$, and width (error) 
$\varepsilon$. 
Unambiguous estimation of the MPB by means of sampling is possible when the size of the sample is sufficiently large 
to ensure a rather precise estimation of $P_{\rm max}$. Let us assume that we 
sample ${\cal N}$ times from the coarse-grained boson distribution, 
with the same input configuration ${\bm \psi}$, and  we record the coarse-grained data 
\be
a_1, a_2, \ldots, a_{\cal N}, 
\label{sup:eq1}
\ee
where $a_j\in \Int_d$ is the label of the $j$th recorded bin. 
The probability $P(\beta|{\bm \psi};\hat{\cal U})$ is then approximated by the frequency of 
occurrence of the label $\beta$ in the original sample (\ref{sup:eq1}), which is given by 
$p_\beta:=n_\beta/{\cal N}$, where $n_\beta$ is the number of occurrences. The bin with the largest 
frequency of occurrence $p_{\max}:=\max_\beta\{p_\beta\}$, is the empirical MPB, i.e., our guess for the 
actual MPB. 

Clearly,  $p_{\max}$ is a random variable, i.e., it varies from sample to sample, and its distribution depends on the boson distribution we sample from. To obtain a CI on 
$P_{\rm max}$, one needs to know how much the distribution of $p_{\max}$ varies around $P_{\rm max}$. 
For  perfect boson sampling it has been shown  that one needs a  sample size of at least  
\bea
{\cal N}^{(0)} =\frac{12d}{\varepsilon^2}\ln(2\gamma^{-1}),
\label{sup:sample_size}
\eea
to ensure
\bea
{\rm Pr}[p_{\rm max}-\varepsilon/2<P_{\rm max}<p_{\rm max}+\varepsilon/2]>1-\gamma,
\eea
for $\varepsilon< 1/d\ll 1$ and $\gamma\ll 1$ \cite{NikBroPRA16}. 
This is the $100(1-\gamma)\%$ CI for $P_{\rm max}$, and the sample size ${\cal N}^{(0)}$ quantifies the cost for the establishment of the CI. 
 The growth of the sample size is basically determined by the relative error  and the number of bins, whereas the dependence on $\gamma$ is 
only logarithmic. Equation (\ref{sup:sample_size}) shows that attaining high accuracy is far more costly 
than getting high confidence. 
As mentioned above, throughout this work we assume that $d$ scales polynomially with $\{M,N\}$,  and thus ${\cal N}^{(0)} $  also scales polynomially with $\{M,N\}$.

Equation (\ref{sup:sample_size}) relies on  the Chernoff bound, and 
it is applicable to any combination of parameters $\{M,N,d\}$. 
At various stages of its derivation we have adopted the worst-case 
scenario  \cite{NikBroPRA16}, which suggests that in practice the same CI could be attained for smaller sample sizes i.e., smaller costs. 
Of course, the main quantity of interest in the present work is the label of the MPB of the coarse-grained distribution, and Eq. (\ref{sup:sample_size}) is of relevance  if one tries to find the MPB, through the estimation of the maximum probability. However, this is not the only available approach. Algorithm \ref{tab:algorithm1}, which is discussed in Sec. \ref{secIII}B,  relies on the bootstrap technique and allows for a direct estimation of the MPB, without the need for an explicit estimation of the maximum probability $P_{\max}$. For the sake of completeness, and in order to gain further insight into the performance of algorithm \ref{tab:algorithm1}, in Sec. \ref{secIIIa} we also discuss the performance of the bootstrap technique with respect to the estimation of $P_{\max}$, although this estimation is not explicitly involved in algorithm \ref{tab:algorithm1}. 


\section{Estimation of the most-probable bin through bootstrap analysis of boson sampling data}
\label{secIII}

Bootstrap is a standard statistical technique \cite{book4,book5}, which allows one to  assign 
measures of accuracy (e.g., confidence intervals) to a statistic estimated on some random sample data. 
Such a statistic may be the sample mean, the sample variance, etc, and the main advantage of the 
bootstrap technique is that it does not require any prior knowledge on the underlying distribution. 
In the present work we employ the bootstrap technique to assign confidence interval to  the maximum 
probability of the coarse-grained boson distribution, and to make an educated guess for the MPB, based 
solely on the data that have been obtained from coarse-grained boson sampling (CGBS). 
The procedure is as follows. 

Using a standard computer, we re-sample from the original sample data (\ref{sup:eq1}) with substitution, to produce an empirical bootstrap sample of size ${\cal N}$. The same procedure is 
repeated ${\cal M}$ times resulting in ${\cal M}$ independent bootstrap samples:  
\begin{subequations}
 \begin{eqnarray}
\label{sup:eq2}
&&\textrm{bootstrap sample 1}:\quad b_{1,1}, b_{1,2}, \ldots,  b_{1,{\cal N}}, \\
&&\textrm{bootstrap sample 2}:\quad b_{2,1}, b_{2,2}, \ldots,  b_{2,{\cal N}}, \\
&&\quad\quad\quad\quad\vdots\nonumber\\
&&\textrm{bootstrap sample } {\cal M}:\quad b_{{\cal M},1}, b_{{\cal M},2}, \ldots,  b_{{\cal M},{\cal N}}.
\end{eqnarray} \end{subequations}
According to the bootstrap principle, the variation of  the difference $\Delta := p_{\rm max}-P_{\rm max}$ is well approximated by the variation of  $\Delta^* := p_{\rm max}^{*}-p_{\rm max}$, where $p_{\rm max}^{*}$ is 
evaluated from an empirical bootstrap sample. That is, $p_{\rm max}^{*}$ is the maximum recorded  frequency of occurrence of a bin in a single bootstrap sample, and let $\mu^*$ denote the label of the corresponding bin. 
The power of bootstrap relies on the fact that 
$\Delta^*$ is obtained by re-sampling directly from the original data (\ref{sup:eq1}) with substitution, 
and the procedure does not involve the calculation of any permanents. This is a straightforward computational 
task, and by means of a standard computer, one can obtain as many bootstrap samples as necessary, to 
estimate $\Delta^*$ to high precision.  

For each one of the ${\cal M}$ bootstrap samples we compute  $\Delta^* := p_{\rm max}^{*}-p_{\rm max}$, and the results are sorted in ascending order (from smallest to biggest). Moreover, we keep track of the 
most-frequent bin in each bootstrap sample, thereby obtaining a sequence of most-frequent bins 
\bea
{\mathbb M}=\{\mu_1^{*}, \mu_2^{*},\ldots,\mu_{\cal M}^{*}\},
\label{sup:mpb_seq}
\eea
where $\mu_j^{*}\in\Int_d$ is the most-frequent bin in the $j$th bootstrap sample. 
By definition, the $100(1-\gamma)\%$ confidence interval for $P_{\rm max}$ is 
\bea
{\rm Pr}[p_{\rm max} - \Delta_{(1-\gamma/2)}^{*}\leq P_{\rm max}\leq p_{\rm max} -\Delta_{(\gamma/2)}^{*}]=1-\gamma,\nonumber\\
\eea
where $\Delta_{(\zeta)}^*$ denotes the $(100\times\zeta)$th percentile of the ordered list of bootstrap values for 
$\Delta^*$. The width of the CI is given by 
$\varepsilon= \Delta_{(1-\gamma/2)}^{*}-\Delta_{(\gamma/2)}^{*}$, 
and quantifies the error in the estimation of $P_{\max}$. 
An educated guess about the label of the actual MPB, may rely on the sequence of most-frequent bins (\ref{sup:mpb_seq}).  Let  $\Omega_\beta$ denote the frequency of appearance of the bin $\beta$ in the sequence 
(\ref{sup:mpb_seq}), and let $\Omega_{\max}:=\max_\beta\{\Omega_\beta\}$. 
The bin that appears more often in the sequence (\ref{sup:mpb_seq}), is most likely to be the actual MPB. 
That is, the label of the empirical MPB is $\tilde{\mu}$ such that $\Omega_{\tilde{\mu}} = \Omega_{\max}$, and 
may or may not coincide with the label of the actual MPB $\mu$. This point is clarified below.

\begin{figure}
\centerline{\includegraphics[scale=0.45]{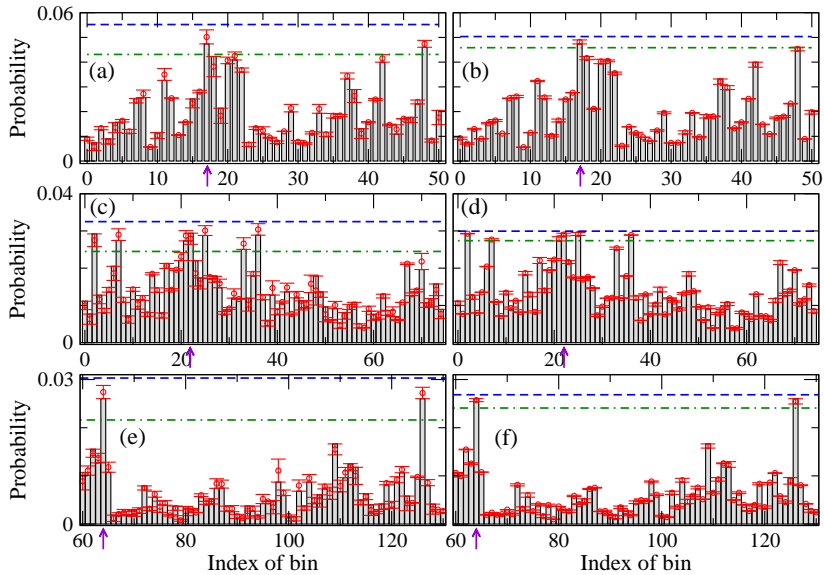}}
\caption{ 
(Color online) Coarse-grained boson sampling for $M=26$, $N=3$. 
The histograms show the exact coarse-grained distribution we sample from. 
The (red) circles with the error bars show the $99.9\%$  CIs for the probabilities of the various bins, while   
the horizontal dashed blue and dot-dashed green lines define the  $99.9\%$  CIs for the maximum probability.  
The CIs have been obtained by means of the bootstrap technique for ${\cal M} = 10^4$.  
The plots on the same row pertain to the same input configuration and the same number of bins, but 
different sample sizes (i.e., ${\cal N} = 10^4$ on the left and ${\cal N} = 10^5$ on the right).  
The arrows point at the actual MPB in each case. 
Number of bins: $d=51$ (a,b); $d=75$ (c,d); $d=151$ (e,f). 
Index of input configuration (following the ordering of Ref. \cite{NikBroPRA16}): 
16 (a,b); 753 (c,d); 16 (e,f).   
}
\label{fig1}
\end{figure}

\subsection{Performance of the bootsrtap technique}
\label{secIIIa}

The performance of the bootstrap technique in the framework of boson sampling has been analysed 
by means of extensive simulations.
Our simulations have been restricted 
to values of $M$ and $N$, for which the problem under consideration is within reach of our computational capabilities \cite{NikBroPRA16}. For a fixed randomly chosen unitary, we were able to construct exactly (up to numerical uncertainty) the entire  boson distribution, for any input configuration and any number of bins. Having the boson distribution, we were able to simulate CGBS for any number of bins, and to 
study the sample sizes required for attaining high-confidence intervals to the maximum probability of the 
coarse-grained distribution, and for identifying the MPB. 

As an example, consider the case of $M=26$, $N=3$ for fixed unitary,  two 
different input configurations ${\bm \psi}$, and three different numbers of bins (see Fig. \ref{fig1}). 
In the first choice of ${\bm \psi}$ and $d$ [see histograms in Figs. \ref{fig1}(a-b)], 
the coarse-grained distribution exhibits a clear dominant peak (at the 17th bin), 
which differs from the second largest peak (at the 48th bin) by $\delta_1\simeq 1.3\times 10^{-3}$. 
By obtaining a sample of size ${\cal N}=10^4$ and ${\cal M} = 10^4$ bootstrap samples, one can 
establish a $99.9\%$ CI for the maximum of the distribution, which has a width $\sim 10^{-2}$ 
[see Fig. \ref{fig1}(a)].  Increasing the 
sample size to ${\cal N}=10^5$, the width of the CI decreases to about $4.5\times 10^{-3}$, and 
the overlap between the CI for the maximum probability and the CI for the second largest peak, 
decreases considerably [see Fig. \ref{fig1}(b)].  
According to Eq. (\ref{sup:sample_size}), without the use of the bootstrap method, 
one has to sample at least $2\times 10^8$ times from the coarse-grained boson distribution,
in order to attain a 99.9\% CI of the same width. In Figs. \ref{fig1}(c-d) we have chosen another 
input configuration and a  larger number of bins.  
The probabilities for the four dominant peaks (at the 2nd, the 22nd, the 25th and the 36th bins) lie within an interval $\delta_{2}\simeq 6\times 10^{-4}$. 
Applying the bootstrap method for ${\cal N} = 10^4$ and  ${\cal M} = 10^4$, we obtain a $99.9\%$ CI for 
the maximum with width $\sim 0.008$, which has a large overlap with the CIs of various 
bins [see Fig. \ref{fig1}(c)]. 
For ${\cal N} = 10^5$, the width of the CI for the maximum decreases to about $0.002$ [see Fig. \ref{fig1}(d)]. According to the Chernoff bound, without the bootstrap technique 
a 99.9\% CI of such width  cannot be attained for sample sizes below $10^{9}$. 
An analogous behavior is depicted in Figs. \ref{fig1}(e,f), where the main two dominant peaks of the 
coarse-grained distribution differ only by $\delta_3\approx 2\times 10^{-5}$.

\begin{figure}
\centerline{\includegraphics[scale=0.5]{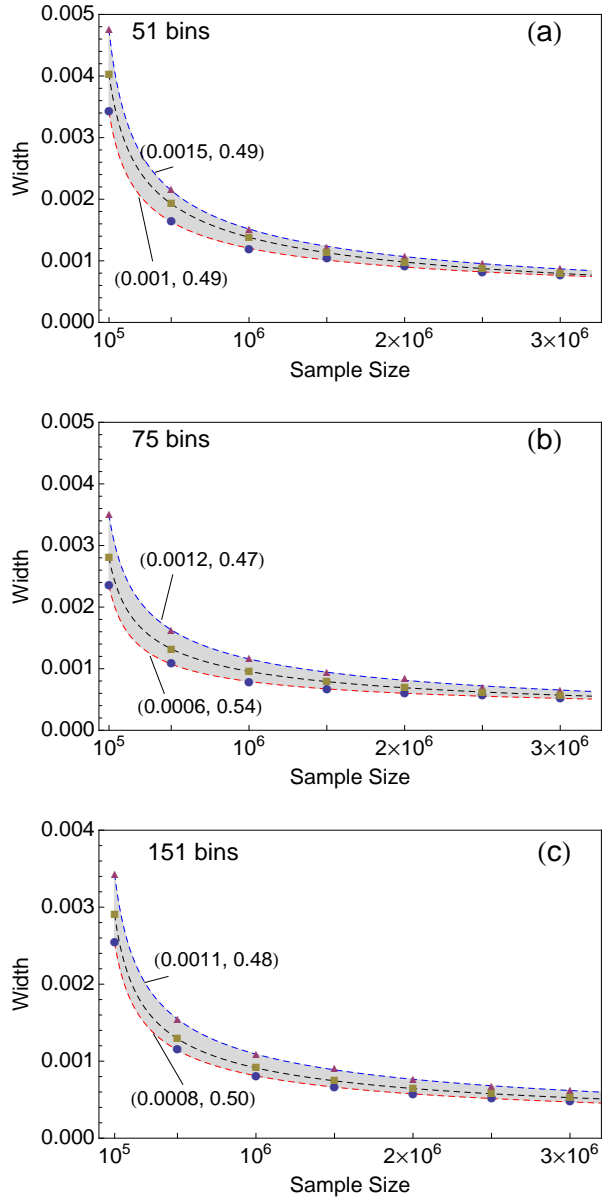}}
\caption{
(Color online) 
The width of the CI for the maximum probability of the coarse-grained distributions of Fig. \ref{fig1}, is plotted for different 
sample sizes.  The symbols show the recorded maximum,  minimum and  central values of the width, in 500 independent realizations. The dashed lines are fits to the numerical 
data of the form $\alpha{\cal N}^{-\beta}$. The estimated values of  
$(\alpha,\beta)$ for the two outermost curves are also shown. 
The parameters used in the three figures are the same as in the rows of Fig. \ref{fig1}.   }
\label{fig2}
\end{figure}

The sample sizes considered in Fig. \ref{fig1} are not sufficiently large to ensure CIs 
for the maximum probability, which are narrower than the difference of the 
probabilities 
of the two dominant peaks in the depicted  coarse-grained distributions.  
However, from the discussion above it is clear  that using the bootstrap 
technique one can 
reduce considerably the sample sizes required for attaining a CI of a certain width, 
relative to the required sample sizes without the  use of the technique (as predicted 
by the Chernoff bound).  
It is also worth emphasizing that Fig. \ref{fig1} is for a single realization, and a CI is 
a random interval, 
whose center and width vary from realization to realization. In any case, 
as depicted in Fig. \ref{fig2},  the width for the CI of the maximum decreases 
with increasing sample size as $\alpha{\cal N}^{-\beta}$, 
for positive $\alpha, \beta<1$. The precise values of $(\alpha,\beta)$ depend on 
the number of bins, as well as on the details of the coarse-grained distribution.

\begin{figure}[t]
\centerline{\includegraphics[scale=0.3]{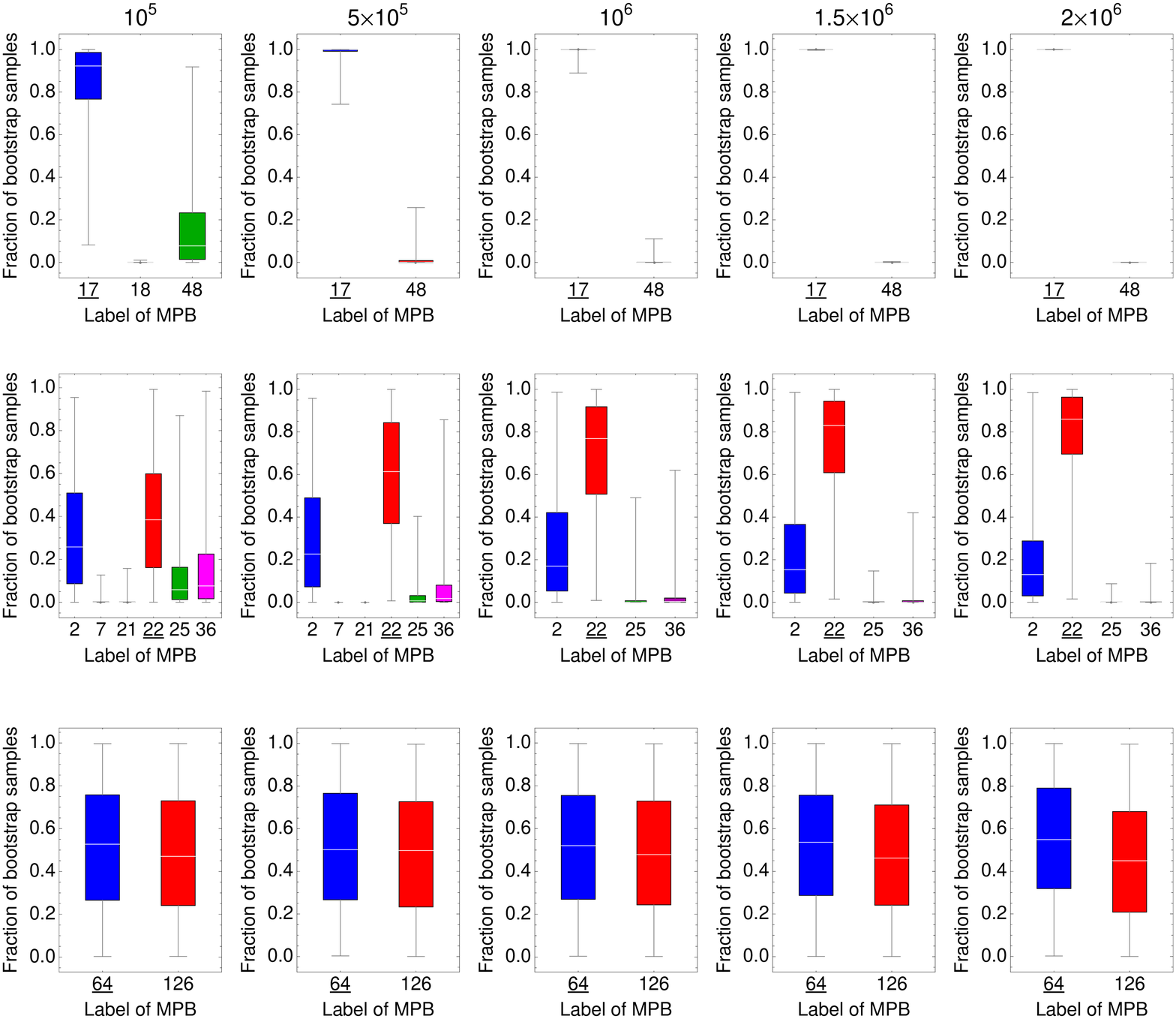}}
\caption{(Color online) Box and whisker diagrams for the data on the fraction of bootstraps that result in the 
same most-frequent bin. 
The diagrams on the same row pertain to the same input configuration, the same number of bins, 
and different values of the same sample size ${\cal N}$ (shown on the top). 
Each diagram has been obtained by means of 500 independent realizations. 
For each realization the data were analysed by means of ${\cal M} = 10^4$ bootstraps, and we recorded the 
fraction of bootstraps that resulted in the same most frequent bin.  
For each box the lower and upper bounds refer to the first and the third quartiles, and the horizontal (white line) the median. The vertical whiskers (error bars) show the extrema of the recorded fractions. For the bins that are 
not included in the diagrams the fraction of bootstrap samples is zero.  The label of the 
actual MPB is shown in bold face and underlined. Other parameters as in the rows of Fig. \ref{fig2}. 
}
\label{fig3}
\end{figure}

Analogous variations are present in the frequency of occurrence $\Omega_\beta$ 
of the bin $\beta$ in set ${\mathbb M}$, which is the main quantity of interest.  
Such variations are shown in Fig. \ref{fig3} for the same parameters as in  
Fig. \ref{fig1}, and  various sample sizes that go beyond the sample sizes in Fig. \ref{fig1}. 
The diagrams in the first row correspond to the distribution (bar chart) depicted in Figs. \ref{fig1}(a,b), 
which exhibits one clear dominant peak (at the 17th bin) that differs by $\delta_1$ from 
the second largest peak (at the 48th bin). As depicted in the graphs of the top row of Fig. \ref{fig3},  
the overlap between the boxes and the 
whiskers vanish very quickly as one increases the sample size, while keeping ${\cal M}$ constant.
Consider now the distributions depicted in Figs. \ref{fig1}(c,d), where the probability 
of the MPB  (the 25th bin),  differs from the probabilities of the 2nd, the  25th and the 36th bins by at most $\delta_2$. Once more we find that the bootstrap method 
identifies correctly all of 
the dominant peaks in the distribution, and the overlap between the various boxes and whiskers decreases 
with increasing sample size. In this case, however, the convergence is a bit slower than 
in the graphs on the top of the same figure, due to the fact that $\delta_2<\delta_1$.  
The situation is fundamentally different in the diagrams at the bottom row of  Fig. \ref{fig3}, which 
correspond to the probability distribution shown in Figs. \ref{fig1}(e,f).  
In this case the two dominant peaks differ by 
$\delta_3\approx 2\times 10^{-5}$, and we find  a 
large overlap of the boxes 
as well as of the whiskers  for sample sizes up to ${\cal N}\simeq 2\times 10^7$. Hence, in this range 
of sample sizes the two peaks are practically indistinguishable, and one can only make a guess for the 
actual MPB, with the probability of a wrong guess being particularly high $(\sim 0.5)$. 

The above findings show clearly that the bootstrap technique is capable of identifying the dominant 
bins in the coarse-grained distribution. Moreover, in the limit of ``large" sample sizes,  
all of the bootstrap samples are expected to return the same most-frequent bin, which coincides with the 
actual MPB of the underlying distribution. The values of the sample size ${\cal N}$ for which this limit  is attained 
depends on the  coarse-grained distribution ${\mathscr P}({\bm \psi};\hat{\cal U})$ we sample from,  and in particular on the distance $\delta$ between the two largest probabilities. The bootstrap technique is capable of 
distinguishing between the MPB and the bin with the second largest probability when the width of the 
CI for the maximum probability satisfies $\varepsilon\lesssim \delta$.  
These findings dictate a strategy for the inference of the MPB from the data of a single realization, which is 
summarized in  algorithm \ref{tab:algorithm1}, and is discussed in the next subsection. 

\subsection{Algorithm for estimation of the most-probable bin in a coarse-grained boson distribution}
\label{secIIIb}

\setcounter{table}{0}

\begin{table}
\renewcommand{\tablename}{ALGORITHM}
\caption{\label{tab:algorithm1}%
Estimation of the MPB of an unknown coarse-grained boson 
distribution ${\mathscr P}({\bm \psi};\hat{\cal U})$, for fixed ${M,N}$ and $d\sim {\rm poly}(M,N)$, by means of  bootstrap analysis. }
\begin{tabular*}{\textwidth}{@{}l*{15}{@{\extracolsep{0pt plus 15pt}}l}}
\hline
{\bf Input}: integers ${\cal M}\gg 1$,  $\delta{\cal N}> {\cal M}$,  $l=1$, and $L>1$. A real positive number $\xi\ll 1$\\ and {\tt STATUS} = {\tt INIT}.\\
\hspace*{0.2cm}1. Sample $\delta{\cal N}$ times from ${\mathscr P}({\bm \psi};\hat{\cal U})$.   \\
\hspace*{0.2cm}2. Analyze together the sample data from all of the $l$ rounds by  means of the\\ \hspace*{0.6cm} bootstrap technique,  to obtain the sequence of most-frequent bins (\ref{sup:mpb_seq}).   \\
\hspace*{0.7cm} (a) If $\Omega_{\max}<1-\xi$ and $l<L$, increase $l$ by 1 and return to step 1. \\
\hspace*{0.7cm} (b) If $\Omega_{\max}<1-\xi$ and $l=L$,  set {\tt STATUS} = {\tt ABORT}. \\
\hspace*{0.7cm}  (c) If $\Omega_{\max}\geq 1-\xi$ and $l\leq L$,  set {\tt STATUS} = {\tt END}, 
and accept $\tilde{\mu}: \Omega_{\tilde{\mu}} = \Omega_{\max}$\\ 
 \hspace*{1.4cm}   as the empirical MPB.   \\
 {\bf Output}:   {\tt STATUS},  and a guess for the label of the MPB $\tilde{\mu}\in\Int_d$, if   {\tt STATUS} = {\tt END}.\\
\hline
\end{tabular*}
\end{table}

\begin{figure}[t]
\centerline{\includegraphics[scale=0.7]{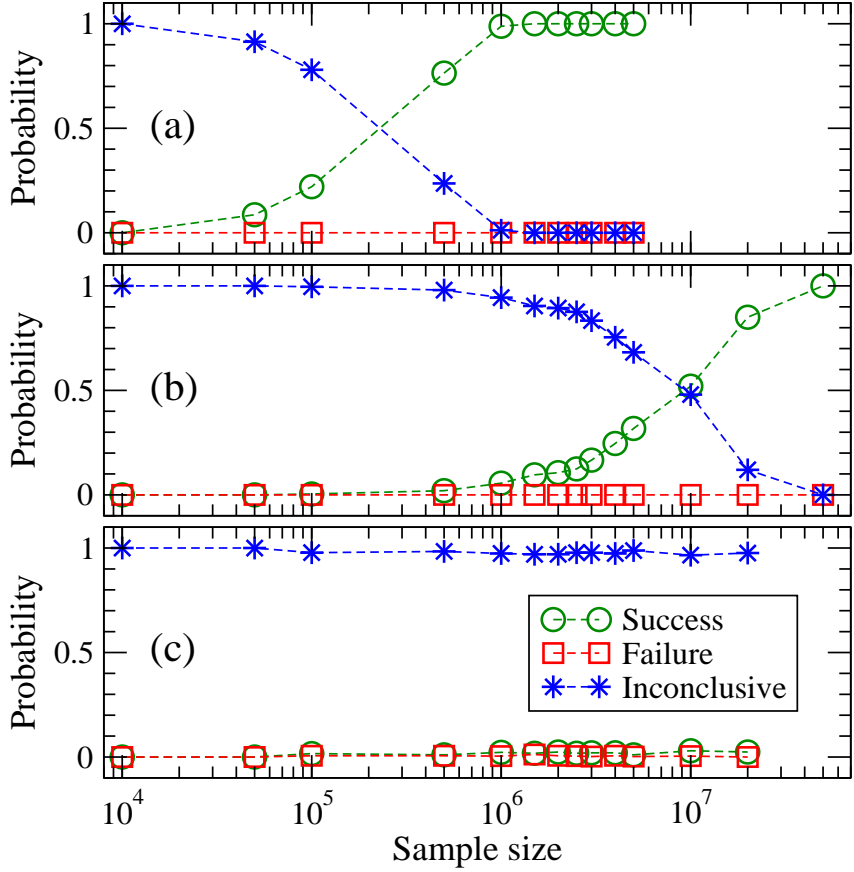}}
\caption{ 
(Color online)
The probabilities of success, failure, and inconclusive result for algorithm \ref{tab:algorithm1}, are plotted as 
functions of the sample size ${\cal N}_{\max}^{({\rm I})}$, for $\xi = 10^{-2}$.  
The parameters used for  each row are the same as for the rows of Figs. \ref{fig2} and \ref{fig3}. 
}
\label{fig4}
\end{figure}

The main idea behind algorithm \ref{tab:algorithm1} is to increase gradually the sample size until the majority of the ${\cal M}$ 
bootstrap samples [at least $100(1-\xi)\%$ of them for $\xi\ll 1$], return the same most-frequent bin. In each round the sample size  is increased by $\delta {\cal N}$, and the algorithm aborts if the predetermined majority has not been attained, 
and the total sample size has reached an upper bound ${\cal N}_{\max}^{({\rm I})}:=L\delta{\cal N}$, which is determined by our computational capabilities. 
Execution of the algorithm \ref{tab:algorithm1}, may result in three different scenarios. 
(a) {\tt STATUS} = {\tt END} and $\tilde{\mu} = \mu$: The algorithm has been successful in identifying the actual MPB. 
(b)  {\tt STATUS} = {\tt END} and $\tilde{\mu} \neq \mu$: The algorithm has failed to identify the actual MPB. 
(c) {\tt STATUS} = {\tt ABORT}: The algorithm has returned an inconclusive result. 
The performance of the algorithm can be quantified in terms of  the corresponding probabilities i.e.,  
 the probability of success $p_{\rm s}$, the probability of failure $p_{\rm f}$, 
and the probability of an inconclusive result  $p_{\rm ?}$. 
These probabilities were estimated numerically, in the framework of independent realizations (see appendix \ref{app2}), and our main 
findings for the cases considered in Figs. \ref{fig2} and \ref{fig3}, are depicted in Fig. \ref{fig4}. 
In all of the cases, for small sample sizes we have $p_{\rm ?} = 1$, $p_{\rm f} = 0$ and $p_{\rm s} = 0$. 
In general, the probability of inconclusive result decreases with increasing sample size, 
whereas the probability of success increases by the same 
amount, and the probability of failure remains practically negligible. How fast  $p_{\rm s}~(p_{?})$ increases (decreases) depends on how close are the  dominant peaks of the coarse-grained distribution. 
By contrast,  the probability of failure in Figs. \ref{fig4}(a-c) is  $p_{\rm f} \lesssim 1.2\times 10^{-2}$, which is of the order of $\xi$ used in our simulations. This was to be expected, because the choice of $\xi$ in  algorithm \ref{tab:algorithm1} determines the 
confidence in choosing the empirical MPB. The smaller $\xi$ is, the higher the confidence becomes. 
In any case, one can 
combine the algorithm  \ref{tab:algorithm1} with standard error-correction techniques \cite{EC-book}, in order to 
ensure negligible values 
for $p_{\rm f}$. For instance, repeating the algorithm $r$ times  (for odd  $r\geq 3$), and 
deciding on the empirical MPB using majority logic decoding, 
one can reduce the probability of failure by $\lceil r \rceil\geq 2$ orders of magnitude.

\begin{figure}[t]
\centerline{\includegraphics[scale=0.45]{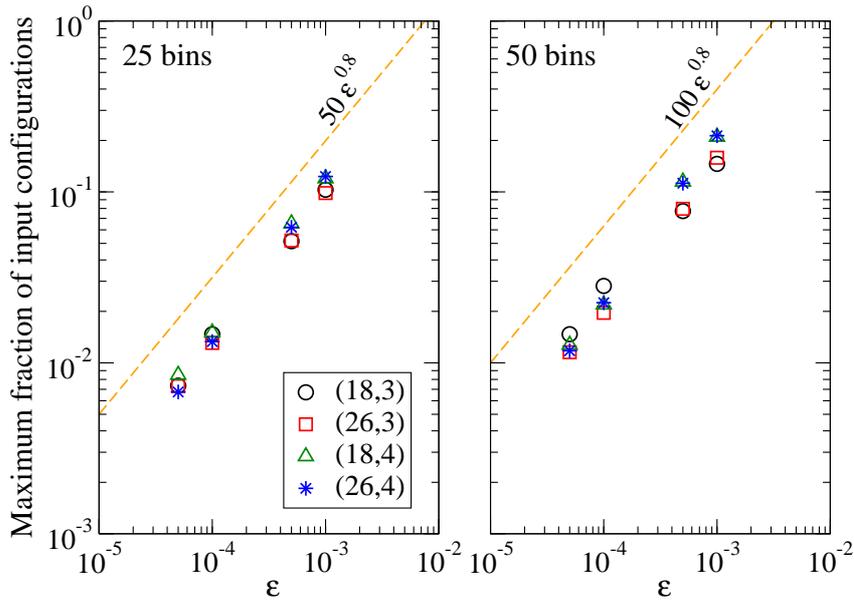}}
\caption{ 
(Color online)
The fraction of input configurations for which the coarse-grained distribution ${\mathscr P}({\bm \psi};\hat{\cal U})$ 
exhibits two or more $\varepsilon$-close dominant peaks, is plotted as function of $\varepsilon$, for two different  values of $d$, and various values of $(M,N)$. The maximum has been obtained over 100  random unitaries, and the dashed curve shows an upper bound given by 
$2d\varepsilon^{0.8}$. 
}
\label{fig5}
\end{figure}

Although not  present in algorithm \ref{tab:algorithm1}, 
the width of the underlying CI for the maximum probability at the end of the $L$  rounds determines the accuracy 
in the evaluation of the maximum probability, and it can help us to understand  deeper the performance of the algorithm. Based on the analysis of Sec. \ref{secIIIa}, 
algorithm \ref{tab:algorithm1} is expected to abort when the 
coarse-grained distribution ${\mathscr P}({\bm \psi};\hat{\cal U})$ exhibits two or more nearly degenerate 
dominant peaks i.e., when $\delta\lesssim \varepsilon$, 
where $\delta$ is the difference of the two largest probabilities in 
${\mathscr P}({\bm \psi};\hat{\cal U})$,  
and  $\varepsilon$ is the width of the CI that has been attained by the algorithm. 
For a given $d$, the existence of such degenerate dominant peaks 
(to be referred to hereafter as ${\varepsilon}$-close dominant peaks) depends on the coarse-grained distribution  we 
sample from i.e., on the input configuration  ${\bm \psi}$ and on the unitary $\hat{\cal{\cal U}}$. 
In an effort to estimate how frequent is the presence of ${\varepsilon}$-close dominant peaks in coarse-grained 
boson distributions, we analysed all of the distributions for various combinations of $\{d,M,N\}$, and 100 Haar 
random unitaries. For each combination of $\{d,M,N,\hat{\cal U}\}$ and a given value of $\varepsilon$, 
we estimated the fraction of input configurations for which the corresponding coarse-grained distributions exhibit 
two or more ${\varepsilon}$-close dominant peaks. As expected, for each combination of $\{d,M,N\}$, 
the fraction was found to vary from unitary to unitary, and in Fig. \ref{fig5} we plot the maximum 
recorded fraction $q_{\max}$ for different values of $\varepsilon$. 
Clearly, $q_{\max}$ decreases with $\varepsilon$, and it is 
bounded from above by $2d{\varepsilon}^{0.8}$. 
Analogous behavior has been found for all of the combinations of parameters we studied, which suggests that this 
is a universal bound. It is also worth emphasizing that for the particular choices of $(M,N)$ shown in Fig. \ref{fig5},  
the size of the input space $|{\mathbb S}|$ ranges from about 800 to about 15000. Yet, this broad range is not 
reflected in the corresponding numerical estimates for $q_{\max}$ at a fixed  $\varepsilon$, which remains  
practically constant while varying $(M,N)$. In other words, we find a very weak dependence of  
$q_{\max}$ on $(M,N)$, and thus on $|{\mathbb S}|$, which suggests that the maximum probability for algorithm \ref{tab:algorithm1} to abort is not expected to increase with increasing $|{\mathbb S}|$.

The last issue that must be clarified before closing this section 
pertains to the maximum number of rounds $L$ in algorithm \ref{tab:algorithm1}, 
which essentially determines the maximum allowed sample size 
${\cal N}_{\max}^{({\rm I})} = L\delta{\cal N}$ before the algorithm aborts. Clearly, 
$L$ should depend only on known quantities, and not on the details of the coarse-grained distribution, which are not {\em a priori} known. 
Based on the aforementioned findings, it is desirable for ${\cal N}_{\max}^{({\rm I})}$  to be sufficiently large so that 
\bea
2d{\varepsilon}^{0.8}\lesssim 0.1. 
\label{cond_qmax}
\eea
In this way one ensures that, given a randomly chosen input configuration,  the algorithm will abort with probability $\lesssim 0.1$. 
In algorithm \ref{tab:algorithm1} the estimation of the MPB is not  
achieved through the estimation of the maximum probability, and 
thus  the CI for $P_{\max}$, is not involved in the process. 
However, the results of the previous subsection show that the 
sample sizes required for assigning a high-confidence interval to 
the maximum probability by means of the bootstrap technique, 
do not exceed the sample size predicted by the Chernoff bound. 
The latter may serve as a benchmark for the 
number of rounds $L$ in algorithm \ref{tab:algorithm1}. 
In particular, inequality (\ref{cond_qmax}) together with 
Eq.  (\ref{sup:sample_size}) for low uncertainty 
$\gamma = 5\times 10^{-4}$, dictates  
\bea
{\cal N}_{\max}^{({\rm I})}\gtrsim 1.8 \times 10^5 d^{7/2}.  
\label{L:est}
\eea 
Note that in  
Figs. \ref{fig4}(a) and \ref{fig4}(b), the probability of success in guessing the MPB by means of algorithm \ref{tab:algorithm1} has converged to $1$, for an overall sample size, which is considerably  smaller than the one predicted by the r.h.s of inequality (\ref{L:est}). Analogous behavior has been found for all of the combinations of parameters that 
we have investigated in our simulations (see  appendix \ref{app3} for additional results). 
In view of the polynomial scaling of $d$ with $\{M,N\}$, 
these findings show that  algorithm \ref{tab:algorithm1} is expected to  succeed or to abort for an overall sample size  that  scales polynomially with $\{M,N\}$. Of course, an alternative approach is to let $L$ (and thus ${\cal N}_{\max}^{({\rm I})}$) be determined by the computational capabilities, as mentioned above.

\begin{figure}[t]
\centerline{\includegraphics[scale=0.45]{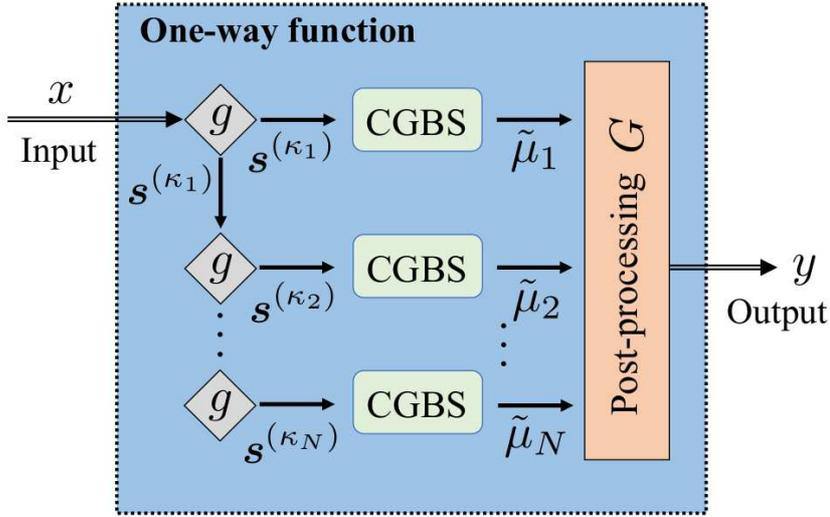}}
\caption{
(Color online) 
The proposed OWF takes as input an integer $x$, chosen at random from a uniform distribution over $\Int_{|{\mathbb S}|}$. Sequential coarse-grained BS (CGBS) is performed  for $N$ boson configurations ${\bm s}^{(\kappa_1)}, {\bm s}^{(\kappa_2)},\ldots$, which are generated from $x$ via a publicly known classical algorithm $g$. The sequence of most frequent bins $\tilde{\mu}_1,\tilde{\mu}_2, \ldots$ is processed though a publicly known algorithm $G$ to yield the output $y\in\Int_{|{\mathbb S}|}$.  }
\label{fig6}
\end{figure}


\section{One-way Function}
\label{secIV}

We assume a publicly known fixed unitary $\hat{\cal U}$, and a publicly known algorithm \ref{tab:algorithm1} with parameters $\{M,N,d\}$. 
 A schematic representation  of the proposed OWF $y={\cal F}(x)$ is given 
in Fig. \ref{fig6}, and a detailed description is given in algorithm \ref{tab:algorithm2}. The input $x$ and the output $y$ of the OWF are 
integers in the set $\Int_{|\mathbb S|}$. The output $y$ basically  determines the position of boson configuration ${\bm \varphi}$ in ${\mathbb S}$, which is generated by $N$ iterations of the CGBS, with seeds that are determined by the input $x$. As discussed in Sec. \ref{secII}, 
there is a unique $y$ in $\Int_{|\mathbb S|}$ such that ${\bm \varphi} = {\bm s}^{(y)}$, with ${\bm s}^{(y)}$ in ${\mathbb S}$.

Algorithm \ref{tab:algorithm1} is invoked in step 1(b) of algorithm 
\ref{tab:algorithm2}.  As discussed above, algorithm \ref{tab:algorithm1} will fail 
to return a conclusive answer about the label of the MPB of the distribution 
${\mathscr P}(\kappa_j;\hat{\cal U})$ [for the sake of simplicity from now on we 
write ${\mathscr P}(\kappa_j;\hat{\cal U})$ instead of 
${\mathscr P}({\bm s}^{(\kappa_j)};\hat{\cal U})$], when there are two or more 
${\varepsilon}$-close maxima  in the distribution.
Given that $x$ is chosen at random, and the details of the distribution are not {\em a priori} known, 
one cannot predict in advance whether algorithm \ref{tab:algorithm1}  will abort or not. 
Hence, when algorithm \ref{tab:algorithm1} aborts,   step 1(b) of algorithm \ref{tab:algorithm2} 
is repeated for a new coarse-grained distribution pertaining to the next input configuration.  Our simulations 
suggest that it is highly unlikely for the coarse-gained distributions of successive input configurations to exhibit 
nearly-degenerate maxima. As long as $L$ is sufficiently large so that $2d\varepsilon^{0.8}\ll 1$, 
repetition of step 1(b) is expected to be a rare event, rather than a regular necessity.

\begin{table}
\renewcommand{\tablename}{ALGORITHM}
\caption{\label{tab:algorithm2}%
One-way function ${\cal F}$, based on the quest for the MPB of a coarse-grained boson distributions 
${\mathscr P}(\bullet;\hat{\cal U})$, for fixed ${M,N}$ and $d\sim {\rm poly}(M,N)$.
}
\begin{tabular*}{\textwidth}{@{}l*{15}{@{\extracolsep{0pt plus 15pt}}l}}
\hline
{\bf Input}: integer $x$ chosen at random from uniform distribution over $\Int_{|{\mathbb S}|}$.\\
\hspace*{0.2cm} 1.  For $j\in[1,N]$:    \\
\hspace*{0.7cm} (a) Algorithm $g$. Estimate  
$\kappa_j = [\kappa_{j-1} + {\rm int}(|{\mathbb S}|\times |\sin(j-1)|)]~{\rm mod}(|{\mathbb S}|)$,\\
\hspace*{1.4cm}  where $\kappa_0 := x$.\\
\hspace*{0.7cm}  (b) Apply algorithm \ref{tab:algorithm1}  to obtain a guess for the MPB of the coarse-grained boson \\
\hspace*{1.4cm} distribution ${\mathscr P}(\kappa_j;\hat{\cal U})$, through the most-frequent bin $\tilde{\mu}_j$. \\
\hspace*{1.4cm} If the algorithm aborts, repeat the step 
for $(\kappa_j+1)~{\rm mod} (|{\mathbb S}|)$. \\
\hspace*{0.2cm} 2. Post-processing $(G)$ of 
$\tilde{{\bm \mu}} = \{\tilde{\mu}_1,\tilde{\mu}_2,\ldots,\tilde{\mu}_N\}$.\\
\hspace*{0.7cm}  (a) Write  the labels of the free ports from $0$ through $M-1$, in ascending order.\\
\hspace*{0.7cm}  (b) Counting from the low end, strike out the $m-$th label not yet struck out, \\
\hspace*{1.4cm}  where 
$m:=\tilde{\mu}_j~{\rm mod}[M_f^{(j)}]$,
 and $M_f^{(j)}$ is the number of free ports that are \\
\hspace*{1.4cm}  available to the $j$th boson. 
 Let  the chosen label be $\varphi_j$.\\
\hspace*{0.7cm}  (c) Repeat step (b) for each element of  
$\tilde{{\bm \mu}}$ to obtain ${\bm \varphi}=(\varphi_1,\varphi_2,\ldots,\varphi_N)$. \\
 {\bf Output}:  integer $y\in\Int_{|{\mathbb S}|}$, which refers to the position of the configuration ${\bm \varphi}$ in ${\mathbb S}$.\\
\hline
\end{tabular*}
\end{table}

The classical sub-algorithms $g$ and $G$ entering our OWF are considered to be publicly known. The general 
structure of the OWF remains unchanged irrespective of the sub-algorithms one may choose, but an imprudent 
choice may compromise the security of our OWF. According to our 
studies, the proposed OWF performs rather well for the particular choice of sub-algorithms, given in 
algorithm \ref{tab:algorithm2}. A sub-algorithm similar to $g$ is 
used in the context of MD5 to generate additive constants in each step \cite{book1}.  
The sub-algorithm $G$ used in algorithm \ref{tab:algorithm2} relies on Fisher-Yates shuffling method, 
and a numerical example of it is given in appendix \ref{app1}. 

\subsection{Properties}
\label{secIVa}

\begin{figure}
\centerline{\includegraphics[scale=0.45]{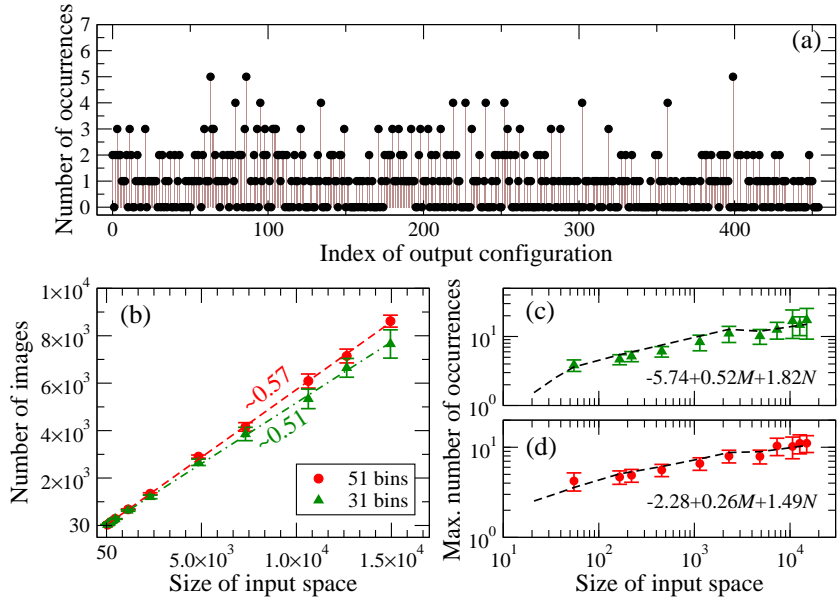}}
\vspace*{0.6cm}
\caption{
(Color online) (a) Number of occurrences (black disks) of the various output configurations, when ${\cal F}$ is evaluated for all possible  inputs (preimages) $x\in\Int_{|{\mathbb S}|}$, in the case of $M=15$, $N=3$, $d=51$. Note the existence of outputs with zero, one or more occurrences. 
(b) The number of images with non-zero occurrences $|{\mathbb Y}|$, for various combinations of 
$\{M,N\}$ (i.e., for various sizes of the input space $|{\mathbb S}|$). The symbols (disks,triangles) refer to 
mean values obtained by averaging over 100 Haar random unitaries, and the error bars denote the corresponding standard deviations.  The dashed lines are linear fits  of the form $|{\mathbb Y}| = \alpha|{\mathbb S}|$ to the numerical data with slopes $\alpha$ shown next to the lines. 
(c) As in (b) for the  maximum number of occurrences $\nu_{\max}$, in the case of $31$ bins. The linear fit to the numerical data is also shown (dashed line). (d) As in (c) for $51$ bins. 
}
\label{fig7}
\end{figure}

The properties of the proposed OWF have been investigated numerically for various combinations of $\{M,N,d\}$.  For a given combination, the above algorithm was applied to evaluate the output (image) $y={\cal F}(x)$ for each one of the possible inputs (pre-images) $x$ in $\Int_{|{\mathbb S}|}$, thereby obtaining the number of occurrences for each  output   $y\in\Int_{|{\mathbb S}|}$. The typical behavior  is depicted in Fig. \ref{fig7}(a), where 
one may notice the existence of outputs with one or more occurrences, as well as the existence of elements $y\in \Int_{|{\mathbb S}|}$ with zero occurrences (i.e., there is no $x\in  \Int_{|{\mathbb S}|}$ such that ${\cal F}(x)=y$). 
These findings were present for all of the combinations of parameters we considered throughout our simulations, and imply  that  the set of all the outputs with one or more occurrences, say  ${\mathbb Y}$, 
is a subset of $\Int_{|{\mathbb S}|}$. 
The largest number of occurrences $\nu_{\max}$ and the size of ${\mathbb Y}$ 
are of particular importance for the security of the OWF, and for a fixed unitary, 
they depend on the combination of the parameters $\{M,N,d\}$. 
As depicted in Fig. \ref{fig7}(b), $|{\mathbb Y}|$ scales linearly with $|{\mathbb S}|$,  
and the slope is determined by $d$. Hence, in view of Eq. (\ref{Ssize:eq}),  
$|{\mathbb Y}|$ scales exponentially with $N$. Moreover, $\nu_{\max}$  depends weakly 
on $d$, and it is well approximated by a linear function 
of $M$ and $N$ [see  Figs. \ref{fig7}(c,d)]. In other words, the largest number of 
occurrences scales polynomially with $(M,N)$ whereas the size of the domain $|{\mathbb S}|$ scales 
exponentially. Given these findings, we can discuss the resources required for the evaluation and 
the inversion of the proposed OWF.

\subsection{Evaluation of the one-way function}
\label{IVb}

Let $T_{\rm c}$  denote the classical  run-time for 
the evaluation of ${\cal F}(x)$ for a given $x$ by means of a classical computer. 
The run-time scales as 
\bea
T_{\rm c} \sim  {\cal N}_{\rm ev}^{\rm (tot)} R_{\rm c} / (\omega n_{\rm cpu}),
\eea
where ${\cal N}_{\rm ev}^{\rm (tot)} $ denotes the total sample size required for the evaluation, 
 $R_{\rm c}$ denotes the number of floating-point operations per sample, $n_{\rm cpu}$ is the number 
 of CPU cores involved in the calculation, while $\omega$ denotes the floating-point operations per second (flops)  and depends on the type of the CPU.
 
The proposed OWF involves $N$ rounds of CGBS and for each round algorithm \ref{tab:algorithm1} is called, which according to the aforementioned discussion, 
is expected to converge for an overall 
sample size  that  scales as $\sim d^{7/2}$. The fastest known classical algorithm for boson sampling 
requires ${\mathscr O}(N2^N)$ operations to produce a single sample \cite{Neville17}, and thus we obtain  
\bea
T_{\rm c} \sim d^{7/2} N^2 2^N  (\omega n_{\rm cpu})^{-1}.   
\eea
Given that $d$  scales polynomially with $(M,N)$, the evaluation of the 
function for 
a given $x\in\Int_{|{\mathbb S}|}$ is only polynomially harder than the evaluation of a permanent of an 
$N\times N$ complex matrix by means of Ryser's algorithm, whose run-time  is 
${\mathscr O}(N2^N)$ \cite{Neville17,NikBroPRA16}. 
In view of recent numerical studies on the computation of permanents \cite{Neville17,NikBroPRA16,Perm16},  evaluation of the proposed OWF for $N\lesssim 40$ is expected to be a feasible task for classical computers. 

\subsection{Inversion of the one-way function}
\label{secIVc}

Consider an adversary who is given $y={\cal F}(x)$ and his task is to find an input 
$\tilde{x}\in{\mathbb S}$, such that $y={\cal F}(\tilde{x})$. 
The adversary is not asked to find the specific input used by the legitimate user in the evaluation of $y$. Any input that satisfies $y={\cal F}(\tilde{x})$ will do. If ${\cal F}$ were bijective, $x$ would be the only pre-image of $y$ under ${\cal F}$. As discussed above, however,  this is not the case for the OWF under consideration. 

We are not aware of any algorithm  (quantum or classical) that can directly invert the proposed OWF i.e., 
to evaluate $\tilde{x}={\cal F}^{-1}(y)$. 
There are at least two main difficulties to the construction of such an algorithm.  
(i) For a fixed unitary, the coarse-grained distribution 
is not {\it a priori} known, and it depends on the randomly chosen input $x$. (ii) The output of the function also depends 
strongly on the input, in the sense that there is only a very small (relative $|{\mathbb S}|$) number of 
inputs that satisfy $y={\cal F}(\tilde{x})$ [see Figs. \ref{fig7}(c,d) and related discussion].  

A standard approach to the inversion of a OWF, is the exhaustive (or brute-force) search,  where the adversary 
tries successively every possible input until he finds the right one. Given that the proposed OWF is not bijective, the question arises is how many inputs an adversary has to randomly select, before there is greater than $(1-\eta)$  chance for one of the chosen configurations to satisfy $y={\cal F}(\tilde{x})$, for some $0<\eta<1$.  The answer to this question  quantifies the work effort that an adversary has to do, if he conducts an exhaustive search. 

Let $\nu$ denote the number of pre-images that satisfy $y={\cal F}(\tilde{x})$, with 
$1< \nu \ll |{\mathbb S}|$, and let $t^\star:=|{\mathbb S}|-\nu$.  
Given that $\nu_{\max}$  scales linearly with $\{M,N\}$ 
and $\nu\leq \nu_{\max}$, by virtue of inequality (\ref{Ssize:eq}) we conclude that 
the growth of $t^\star$  with $\{M,N\}$ is mainly determined by the exponential 
growth of $|{\mathbb S}|$. 
The probability for the adversary  to fail to find a pre-image in $t$ trials is given by 
\begin{equation*}
P_{\rm f}(t)  =
\begin{cases}
\prod_{j=1}^{t} 
 \left  [1-\frac{\nu}{|{\mathbb S}|-(j-1)} \right  ] & \text{if } 1\leq t\leq t^\star,\\
0 & \text{if } t > t^\star.
\end{cases}
\end{equation*}

The lower branch of this function refers to the unlikely, yet possible, scenario for the adversary not to find a preimage of $y$ in $t^\star$ trials. In this case, the adversary will have performed an exponentially large 
number of unsuccessful trials, and the next guess will be certainly among 
the possible preimages of $y$.  We focus now on the case of $1\leq t\leq t^\star$. 

\begin{figure}
\centerline{\includegraphics[scale=0.45]{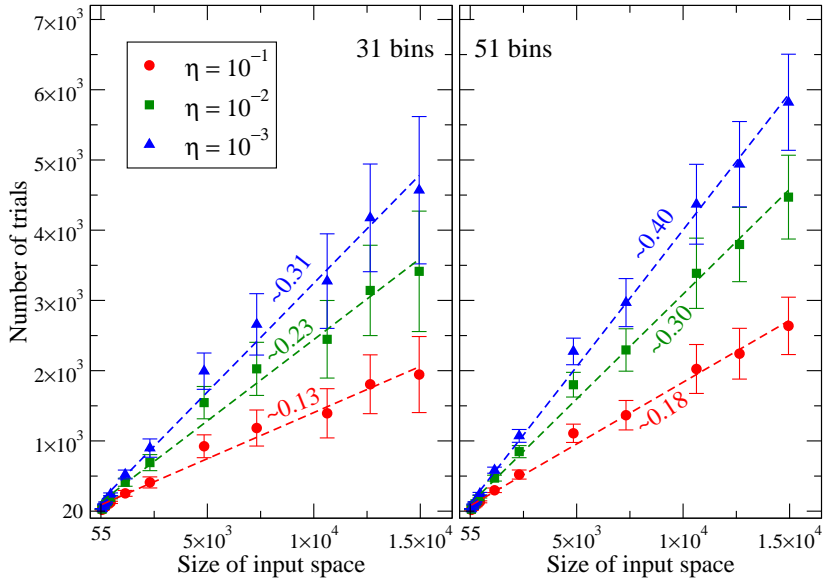}}
\caption{
(Color online) The minimum number of trials  $t_{\rm min}$ given by Eq.  (\ref{tmin:eq}), for various combinations $\{M,N\}$ (i.e., different sizes of the input space $|{\mathbb S}|$) and three different 
values of $\eta$. The symbols (disks, squares, triangles) denote the mean values of $t_{\min}$ over 100 Haar random unitaries, and the error bars denote the corresponding standard deviations.  The dashed lines are linear fits of the form $\nu_{\max} = \alpha|{\mathbb S}|$ to the numerical data, with the slopes $\alpha$ shown next to the lines. Other parameters as in Fig. \ref{fig7}.}
\label{fig8}
\end{figure}

The difference in the brackets decreases with increasing $j$, and thus  
 \bea
P_{\rm f}(t)  &\geq&   
 \left  [1-\frac{\nu}{|{\mathbb S}|-t+1} \right  ]^t\geq \exp\left (-\frac{t\nu}{|{\mathbb S}|-t+1-\nu}\right ). 
\eea
For the second inequality we have used Jensen's inequality $\ln(z)\geq (z-1)/z$, for all $z>0$.  
So, the number of trials required in order for the  probability of failure to be equal to or smaller than 
some positive number $\eta<1$, satisfies 
\bea
t \geq \frac{(|{\mathbb S}|+1-\nu)|\ln(\eta)|}{\nu + |\ln(\eta)|} .  
\eea
For fixed unitary and fixed parameters $\{M,N,d\}$, the number of preimages $\nu$ depend on the given output $y$. Noting, however,  that the l.h.s. of the inequality is a decreasing function of $\nu$, a lower bound on $t$ is given by 
\bea
t_{\min}:=\frac{(|{\mathbb S}|+1-\nu_{\max})|\ln(\eta)|}{\nu_{\max} + |\ln(\eta)|} .  
\label{tmin:eq}
\eea
where $\nu_{\max}$   is the largest number of preimages for the given combination of $\{M,N,d\}$.
As shown in Fig. \ref{fig8},  $t_{\min}$ scales linearly with $|{\mathbb S}|$ and the slope is  determined 
by the number of bins  $d$, and $\eta$. For example, in the case of  $51$ bins, 
the adversary needs at least $0.2|{\mathbb S}|$ and $0.3|{\mathbb S}|$ trials for the probability 
of failure to drop below $\eta=10^{-1}$ and $\eta=10^{-2}$, respectively. 
In view of Eq. (\ref{Ssize:eq}),  these findings suggest that the number of trials and thus the overall work effort 
for an exhaustive search increases exponentially with $N$, irrespective of whether the adversary uses  
a classical or a quantum computer. 

As discussed above, a classical legitimate user is able to evaluate  ${\cal F}(x)$  for a chosen $x$ at run-time  $T_{\rm c} \sim {\rm poly}(N)\times 2^N$.  By contrast, a classical adversary who conducts the exhaustive search using the same classical algorithm will have to perform $t_{\min} \sim N^N$ more operations than the legitimate user. For the sake of concreteness, consider the case of $N= 21$ and $M\geq N^2$. 
The classical run-time of an exhaustive search scales as ${\widetilde T}_{\rm c} \sim (2N)^N  (\omega n_{\rm cpu})^{-1}$, and for $N= 21$ we have  ${\widetilde T}_{\rm c} \sim 10^{34} (\omega n_{\rm cpu})^{-1}$ sec. 
For a supercomputer with $n_{\rm cpu} = 10^4$ and $\omega = 10^{21}$ flops, 
${\widetilde T}_{\rm c}\sim 30$ years. 
Such computing power is not  currently available, and according to Moore's law it is not expected to be 
available before 2030. 
For the time being exhaustive search by means of a BSD is practically impossible for $N\gtrsim 5$, 
because of the very low sampling rates \cite{Neville17}. It has been conjectured that by means of near-term experimental improvements, 20-boson sampling may be possible at a rate of $\sim 130{\rm h}^{-1}$ \cite{ExpBS8}, which cannot compete with the sampling rates of classical computers \cite{Neville17}.  

Given that the proposed OWF relies on a sampling problem and on the evaluation 
of permanents, Shor's algorithm and  its variants cannot be used for its inversion.  Moreover, for the reasons discussed above, the construction of an inversion algorithm is conjectured to be practically impossible. A brute-force quantum search using Grover's algorithm \cite{book3}, is 
the  only currently known attack which can be implemented on a universal fault-tolerant quantum computer. Analogous attacks are applicable to any symmetric or asymmetric cryptosystem, but they are not considered to be a serious threat because the speed-up offered by Grover's algorithm is not as dramatic as Shor's speed-up. More precisely, for the  OWF under consideration Grover's algorithm  is expected to find a preimage with high probability using approximately 
$\sqrt{|{\mathbb S}|/\nu_{\max}}$ evaluations \cite{grover}, as opposed to  
$t_{\min}\sim |{\mathbb S}|/\nu_{\max}$ evaluations required by a classical exhaustive  search algorithm (see Eq. \ref{tmin:eq}). 
Given the exponential growth of $|{\mathbb S}|$ with $N$ and the polynomial dependence 
of $\nu_{\max}$ on $N$, the quantum search algorithm will also need an exponentially large number of operations $(\sim N^{N/2})$.  Hence, the security of the proposed OWF against Grover's algorithm can be increased  by increasing the number of bosons $N$ used in the OWF. It is worth noting that each one of the $\sqrt{|{\mathbb S}|/\nu_{\max}}$ evaluations 
must wait for the previous evaluation to finish. Taking into account this 
limitation various authors believe that it is very likely for Grover's 
improvement to be eliminated in practice by the high cost of qubit operations \cite{pqc2}, thereby making Grover's algorithm useless. 

\subsection{Collision search} 

We have seen that the proposed OWF exhibits collisions i.e., two or more inputs (preimages) $x$ may have the same image $y$ 
under ${\cal F}$. This means that we are essentially dealing with a 
one-way hash function \cite{book1}. The resistance of the proposed OWF to inversion is sufficient for certain cryptographic applications, but there are also applications where one has to ensure its resistance against collision search. For this reason, it is worth discussing here 
the performance of the so-called ``birthday attack" against the 
proposed OWF.
This is also a powerful brute-force attack, which is applicable to any hash function, and in its simplest form proceeds as follows. 
\begin{enumerate}
\item The adversary chooses an input $x$ at random, he calculates 
$y = {\cal F}(x)$, and  stores the pair $(x,y)$  in a database. 
\item The adversary chooses randomly a new input $x^\prime$, and checks whether the new output $y^\prime = {\cal F}(x^\prime)$  already exists in the database.
\item If it does the procedure stops, because a collision has been found. If it does not, then the  new pair $(x^\prime,y^\prime)$ is added  to the database and step 2 is repeated.
\end{enumerate}

The probability of finding at least one collision $P_{\rm suc}^{{\rm (bs)}} $, 
increases with the number of elements in the database $\theta$. 
In particular, in the framework of a generic birthday attack one readily 
obtains for $\theta\ll |{\mathbb W}|$ \cite{stinson:book}
\bea
P_{\rm suc}^{{\rm (bs)}} = 1-\exp\left( -\frac{\theta(\theta-1)}{2|{\mathbb W}|}\right ), 
\label{birthday:eq}
\eea
where $|{\mathbb W}|$ is the total number of images (outputs) under consideration. 
Given that this estimate pertains to a generic model, deviations may be expected  when the attack is applied on a particular hash function. 

For the OWF discussed in the present work, we have seen that there are outputs, with 0, 1, or more preimages, which is not taken into account in the derivation of Eq. (\ref{birthday:eq}). Hence, as a first correction, one may  expect that Eq. (\ref{birthday:eq}) will be valid when the birthday attack is launched against our OWF, but with   some effective size of the output space $|{\mathbb W}| = \sigma |{\mathbb S}|$, where  the constant $\sigma$ depends on $\{M,N,d,\hat{\cal U}\}$.

\begin{figure}
\centerline{\includegraphics[scale=0.45]{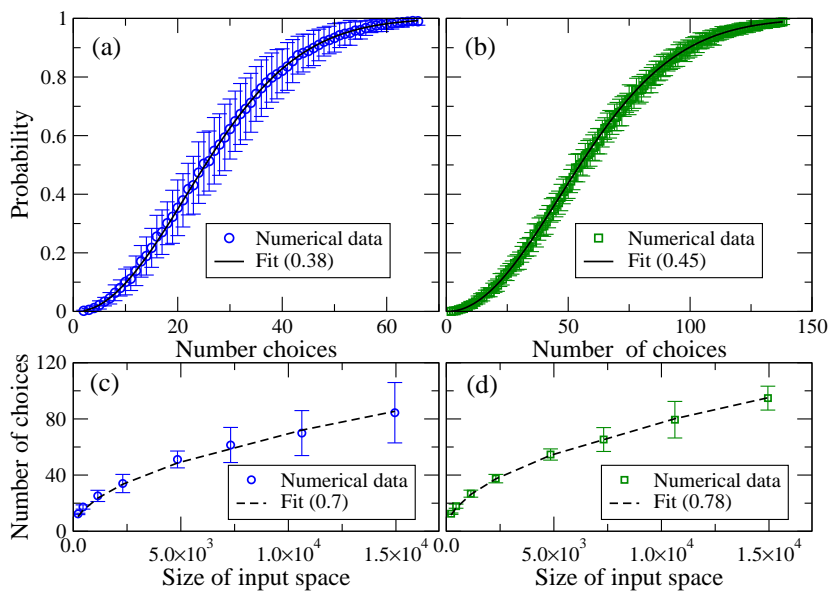}}
\caption{(Color online) Performance of the birthday attack on the OWF. 
(a) The probability of finding a collision as a function of the number of random choices $\theta$, for $M = 20$, $N= 3$, $d=31$.  For each $\theta$, the symbol shows the average over 100 Haar random unitaries, and the error bar denotes the maximum and minimum recorded values. The solid curve is a fit of Eq. (\ref{birthday:eq}) to the numerical data, with $|{\mathbb W}| = \sigma|{\mathbb S}|$ and the value of $\sigma$ shown in the legend. (b) Same as (a) for $M = 20$, $N=4$, $d=51$.  (c,d) The number of random choices $\theta^{*}$ required for $P_{\rm suc}^{\rm(bs)} = 1/2$, is plotted as function of the size of input space $|{\mathbb S}|$, for $d=31$ (c) and $d=51$ (d). Symbols and error bars as in (a,b). The dashed curve is a fit of $\alpha\sqrt{|{\mathbb S}|}$ to the numerical data, with the proportionality constant $\alpha$ shown in the legends.}
\label{fig9}
\end{figure}

In order to confirm this,  and to obtain an estimate of 
the constant $\sigma$, we have simulated numerically the  birthday attack on the OWF under consideration, for various combinations of parameters, and our main findings are summarized in Fig. \ref{fig9}. 
Figures \ref{fig9}(a,b), show clearly that there are small variations with the choice of the unitary, but the increase of $P_{\rm suc}^{{\rm (bs)}}$ with $\theta$ follows the law of Eq. (\ref{birthday:eq}). An estimate for the effective  size of the output space can be obtained by fitting Eq. (\ref{birthday:eq}) to the averaged data. Moreover, in 
Figs. \ref{fig9}(c,d) 
we plot the number of random choices $\theta^*$ required for $P_{\rm suc}^{{\rm (bs)}}=1/2$, as a function of the size of the input space 
$|{\mathbb S}|$. Our results suggest that 
$\theta^* \simeq  \alpha \sqrt{|{\mathbb S}|}$, where the proportionality constant $\alpha$ is determined fully by the desired value of $P_{\rm suc}^{{\rm (bs)}}$ and the number of bins.  Thus, by virtue of Eq. (\ref{Ssize:eq}) and the polynomial scaling of $d$ with 
$N$, we have that $\theta^*$ scales exponentially with $N$ (i.e., $\theta^*\sim N^{N/2}$). Hence, the birthday attack imposes a lower bound on the  number of bosons, so that the function under consideration is secure against collision search with certain computing power. 

A few remarks are  in order before closing this section. (i) Quantum computers are not expected to be more efficient than classical computers in finding collisions. (ii) The resistance of an elgorithm against collision search also implies second-preimage resistance \cite{stinson:book}. (ii) Collision resistance is not important for all of the possible applications of a one-way hash function \cite{martin:book}.


\section{Concluding remarks}
\label{secV}

We have discussed a way of exploiting the problem of BS in the design of a OWF. 
For practical reasons, our simulations 
have been restricted to combinations of $\{M,N\}$ for which, exact 
(up to numerical error) construction of all the possible  boson distributions 
through the evaluation of all of the related permanents, 
was within reach of our computational capabilities. 
This tedious task was necessary for two main reasons. Firstly,  it allowed us 
to investigate  the convergence of our algorithms with respect to the sample sizes 
required for an accurate estimation of the MPB of the coarse-grained distribution. Our findings suggest that convergence can be attained for sample sizes that scale polynomially with $\{M,N\}$, and they are comparable to or smaller than the sample sizes that have been obtained analytically using  the Chernoff bound. 
Secondly, we were able to analyze the properties of the proposed OWF, including 
the presence of collisions, and the size of the image space. 

In view of recent results on classical boson sampling algorithms \cite{Neville17}, our analysis 
suggests that evaluation of the proposed OWF for a given input  can be performed 
on a classical computer for $N\lesssim 40$.  
Yet, exhaustive (brute-force) search, which is the only currently 
available approach to the inversion of the function, requires resources that scale with the number of bosons as $\sim N^N$, irrespective of the computing power of the adversary. 
Hence, brute-force inversion of  the proposed OWF can 
be thwarted by choosing sufficiently large values of $N$ (typically $N\gtrsim 21$). The resources required for successful collision search 
by means of a birthday attack also scale exponentially with $N$, but as $\sim N^{N/2}$. 

For the evaluation of the OWF on a classical computer 
one may employ the algorithm in \cite{Neville17}, which relies on  Metropolised independence sampling.  This algorithm does not need 
any particular software, and the authors in Ref. \cite{Neville17} demonstrate samples of size $10^4$ for $N\simeq 20$ bosons on a standard laptop, with the production of a sample  taking less than 30 minutes. Taking into account the numerical techniques  of Ref. \cite{Perm16}, the algorithm of Ref. \cite{Neville17}  allows for sampling with large number of bosons (up to 50), when implemented on a supercomputer. The implementation of the bootstrap technique is straightforward, and does not require any special software or hardware.

The structure of the OWF is rather general and can accept different 
choices for the binning, the sub-algorithms $g$ and $G$, as well as for the algorithm that is used for the estimation of the MPB.  An interesting question is whether there is an optimal way of choosing the bins in the coarse graining of boson-sampling data for given $\{M,N\}$, and whether the present algorithms can be extended to other clustering techniques \cite{BSpattern}. It is also worth investigating whether the OWF can be modified so that to eliminate the possibility of inconclusive results, while keeping the probability of failure negligible. 
In this case, the evaluation of the OWF will become even more efficient, because it will require smaller sample sizes. 
An extension of the proposed OWF to the case of BS beyond the dilute limit should be possible, provided one 
allows for multiple bosons to occupy the same port. It is not clear, however, whether and how such an extension 
will  affect the properties of the OWF discussed above, and an additional thorough investigation is required in order to 
shed light on these issues. 

The present results suggest that 
the usefulness of boson sampling may not be limited to the proof of quantum supremacy, and pave the way towards cryptographic
applications, which offer computational security against both classical and quantum adversaries.  The design 
of specific protocols goes beyond the scope of the present work, and is a subject of future research.  


\section*{Acknowledgments}
The author thanks S. Aaronson and J. J. Renema for useful comments on the coarse-grained boson sampling, as well as 
T. Garefalakis and T. Brougham, for helpful discussions.  This work was supported by the Deutsche Forschungsgemeinschaft as part of the CRC 1119 CROSSING.

\appendix

\section{Appendix: Example of Fisher-Yates shuffling method}
\label{app1}
To  illustrate the application of Fisher-Yates shuffling method in our context, let us consider the case of $M=10$ modes and $N=4$ bosons. Let the sequence of most-frequent bins that has been obtained at the end of 
step 1 in algorithm \ref{tab:algorithm2}  be $\tilde{{\bm \mu}} = \{3, 7, 6, 9\}$. Initially none of the ports is occupied (round 0). As shown in table \ref{tab1}, in the $j$th round  the $j$th element of the set 
$\tilde{{\bm \mu}} $ dictates the port that will be occupied (shown in bold face), and it is deleted from 
the list of free ports. 

\setcounter{table}{0}

\begin{table*}\caption{An example of the algorithm $G$ used in the proposed OWF, for the case of $M=10$ ports, $N=4$ bosons. }
\label{tab1}
\begin{center}
\begin{tabular}{lcccccc}
\hline
Round & $M_f^{(j)}$ & $\tilde{\mu}_j$  & $m$  &  Free ports & Occupied ports \\
\hline
0 &    &  & &  0, 1, 2, 3, 4, 5, 6, 7, 8, 9 & \\ 
1 & 10  & 3 &  {\bf 3} & 0, 1, 2, {\bf 3}, 4, 5, 6, 7, 8, 9 & {\bf 3} \\ 
2 &  9  & 7 & {\bf 7} & 0, 1, 2, 4, 5, 6, 7, {\bf 8}, 9 & {\bf 3}, {\bf 8}\\
3 &   8 & 6 & {\bf 6} & 0, 1, 2, 4, 5, 6, {\bf 7}, 9 & {\bf 3}, {\bf 8}, {\bf 7}\\
4 &  7  & 9 & {\bf 2}  & 0, 1, {\bf 2}, 4, 5,  6,  9 & {\bf 3}, {\bf 2}, {\bf 8}, {\bf 6}\\ 
\hline
\end{tabular}
\end{center}
\end{table*}

\section{Appendix: Evaluation of probabilities} 
\label{app2}
Algorithm \ref{tab:algorithm1} outputs an educated guess $\tilde{\mu}$ for the label of the MPB. 
When the flag ${\tt STATUS} = {\tt END}$,  there are two different scenarios for the guess i.e., either it is equal to the actual label $\mu$, or it is different. In the former case the algorithm has succeeded, whereas in the latter it has failed. A third possibility is for the algorithm not to converge for the given maximum number of  rounds $L$. In this case the algorithm outputs the flag ${\tt STATUS} = {\tt ABORT}$, which refers to an inconclusive outcome. The probabilities of success, of failure and of inconclusive result have been estimated numerically. 
To this end, for a given set of parameters $\{M,N,d, {\cal M},\xi,L\}$,
 a given unitary $\hat{\cal U}$ and a given input configuration 
 ${\bm \psi}$,  we performed 500 independent realizations of the algorithm \ref{tab:algorithm1}. The probabilities were estimated by the following quotients: 
\bea
&&p_{\rm s} = \frac{\textrm{Number of realizations in which }{\tt STATUS} = {\tt END}\, \textrm{and}\, \tilde{\mu} = \mu}{\textrm{Total number of realizations}},
\\
&&p_{\rm f} = \frac{\textrm{Number of realizations in which }{\tt STATUS} = {\tt END}\, \textrm{and}\, \tilde{\mu} \neq \mu}{\textrm{Total number of realizations}},
\\
&&p_{\rm ?} = \frac{\textrm{Number of realizations in which }{\tt STATUS} = {\tt ABORT}}{\textrm{Total number of realizations}}.
\eea

\begin{figure}
\centerline{\includegraphics[scale=0.45]{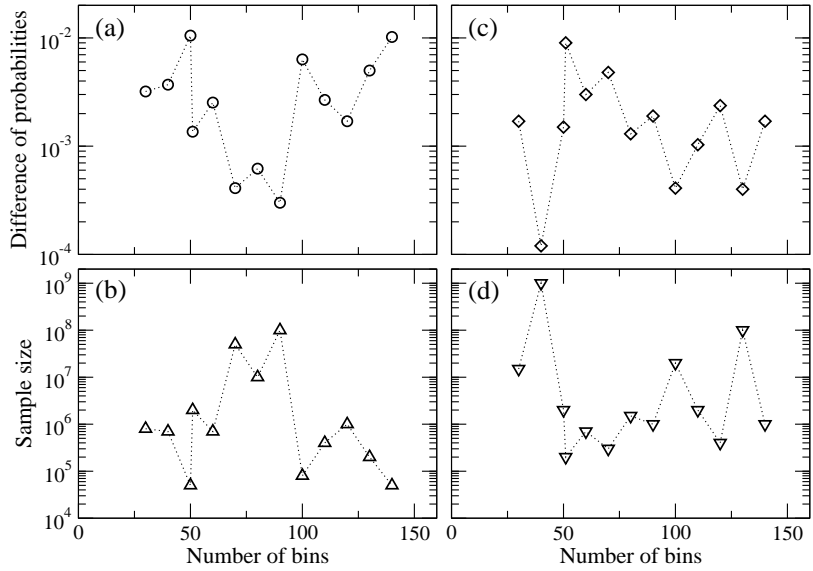}}
\caption{(Color online) Performance of algorithm \ref{tab:algorithm1}, for  two different seeds and a fixed unitary. (a,c) The difference of the largest probabilities in the coarse-grained distribution, as a function of the number of bins. (b,d) The sample size required for algorithm  \ref{tab:algorithm1} to identify the right MPB with probability $p_{\rm s} = 1$, as a function of the number of bins. The input configurations in (a,b) and (c,d) are the same  as in Figs. \ref{fig1}(a,b) and \ref{fig1}(c,d), respectively.  Parameters: $M = 26$ and $N = 3$, ${\cal M} = 10^3$, $\xi = 10^{-2}$, 500 independent realizations.}
\label{fig10}
\end{figure}

\section{Appendix: Performance of algorithm \ref{tab:algorithm1} with respect to  the  numbers of bins} 
\label{app3}

As discussed in Sec. 3, the performance of algorithm \ref{tab:algorithm1} depends on the  difference $\delta$ of the values of the two dominant peaks in the unknown coarse-grained distribution we sample from. In general, for fixed $\{M,N,\hat{\cal U}\}$,  the difference $\delta$ varies with the input configuration ${\bm \psi}$, as well as with the number of bins $d$. Such variations are shown in 
Figs. \ref{fig10}(a,c), for two different input configurations. Moreover, 
as depicted  Figs. \ref{fig10}(b,d), the sample size required for algorithm \ref{tab:algorithm1} to identify the right MPB follows closely the variations of $\delta$ in Figs. \ref{fig10}(a,c), albeit in a very  different scale.  More precisely, it is clear that the sample size tends to increase (decrease)  when $\delta$ decreases (increases). It is also worth noting that the sample sizes depicted in Figs. \ref{fig10}(b,d) are orders of magnitude below the quantity on the r.h.s. of Eq.  (\ref{L:est}), which confirms once more the analysis in Sec. \ref{secIIIb}.

\end{document}